\documentclass[pra,showpacs,twocolumn,amsmath,amssymb,longbibliography]{revtex4-1}

\usepackage{xifthen}% provides \isempty test

\newcommand{\refAppendix}[6]{#1
  \ifthenelse{\isempty{#2}}%
    {}% if #2 is empty
    {\protect\cite{#2}}% if #1 is not empty
    #3\protect\ref{#4}#5#6\xspace
}

\usepackage{forloop,ifthen} % for loops and if statements
\usepackage{multirow} % merge rows and columns
\usepackage{ esint }

\usepackage{amsmath} % need for subequations

    \usepackage{graphicx}
    \usepackage{transparent}
    \usepackage{fancybox}  
    \usepackage{color}
    \usepackage{dcolumn}
    \usepackage{bm}
    \usepackage{xspace}
    \usepackage[colorlinks=true,     linkcolor={red!70!black},
    citecolor={green!50!black}, urlcolor={blue!50!black}, pdfborder={0 0 0}]{hyperref}

\usepackage{calc}

    \newcommand{\VEC}[1]{{\mathbf #1}}
    \newcommand{\CO}[1]{}%SuppressText
    \newcommand{\emphLabel}[1]{\textbf{({#1})}}

\usepackage[usenames,dvipsnames]{xcolor}
\usepackage{ amssymb }

\newcommand{\ps}{phase space\xspace}

\newcommand{\suppress}[1]{}%SuppressText

\newcommand{\emphCaption}[1]{{\bf{#1}}}

\begin{document}

% \title{Quantum Kerr oscillators in \ps: symmetries, Wigner current, shear suppression and
%   special states}

\title{Quantum Kerr oscillators' evolution in \ps: Wigner current, \protect{\\}
  symmetries, shear suppression and special states}

\author{Maxime Oliva and Ole Steuernagel}

\email{Ole.Steuernagel@gmail.com}

\affiliation{School of Physics, Astronomy and Mathematics, University of
Hertfordshire, Hatfield, AL10 9AB, UK}

\date{\today}

\begin{abstract}
  The creation of quantum coherences requires a system to be anharmonic. The simplest such
  continuous one-dimensional quantum system is the Kerr oscillator. It has a number of
  interesting symmetries we derive. Its quantum dynamics is best studied in \ps, using
  Wigner's distribution~$W$ and the associated Wigner \ps current~$\VEC{J}$. Expressions
  for the continuity equation governing its time evolution are derived in terms
  of~$\VEC J$ and it is shown that~$\VEC{J}$ for Kerr oscillators follows circles in
  \ps. Using~$\VEC{J}$ we also show that the evolution's classical shear in \ps is quantum
  suppressed by an effective ``viscosity". Quantifying this shear suppression provides
  measures to contrast classical with quantum evolution and allows us to identify special
  quantum states.
\end{abstract}

% \pacs{03.65.-w, %Quantum mechanics 
%  03.65.Ta %Foundations of quantum mechanics
%}

\maketitle

\section{Introduction \label{sec_intro}}

The formation of quantum coherences is of central importance in the study of quantum
systems and their dynamics.

Here we consider closed one-dimensional Kerr-type oscillators. These are anharmonic
and can therefore create coherences~\cite{Kakofengitis_PRA17}. Additionally, their
dynamics has circular symmetry in \ps. This makes them the simplest continu\-ous system to
create coherences.

In other words, the results reported here apply to regular anharmonic systems (with
Hamiltonians of the form $\hat H = \hat p^2 / 2 m + V(\hat x)$, see~\cite{Oliva_PhysA17} and~\cite{Oliva_Shear_19})
but the Kerr-oscillators' symmetries make them particu\-larly suited to help us understand
aspects of nonclassical effects in quantum dynamics.

Wigner's distribution~$W$~\cite{Wigner_PR32,Hillery_PR84} is the closest quantum
analog~\cite{Zurek_NAT01,Oliva_PhysA17,Hillery_PR84,Leibfried_PT98,Tilma_PRL16} of the
classical \ps distribution~$\rho$. In continuous one-dimensional systems the creation of
quantum coherences is represented by the creation of nega\-tive regions of the Wigner
distribution~\cite{Heller_JCP76,Feynman_NegEssay87,Leibfried_PT98,Zurek_NAT01,Oliva_PhysA17}.
The formation of such nega\-tive regions in the Wigner distribution is easily moni\-tored
numerically.

The evolution of~$W$ is governed by the associated Wigner phase-space current~$\VEC{J}$ (strictly
speaking $\VEC{J}$ is a probability current density). Generally, phase-space-based approaches are
suitable for comparison of quantum with classical
dynamics~\cite{Berry_JPA79,Zurek_NAT01,Oliva_Shear_19}.  Specifically, $\VEC{J}$ allows us
to adopt a geometric
approach~\cite{Ole_PRL13,Kakofengitis_EPJP17,Kakofengitis_PRA17,Oliva_PhysA17,Oliva_Shear_19}
to studying quantum dynamics.

\begin{figure}[t]
  \includegraphics[width=0.49\textwidth]{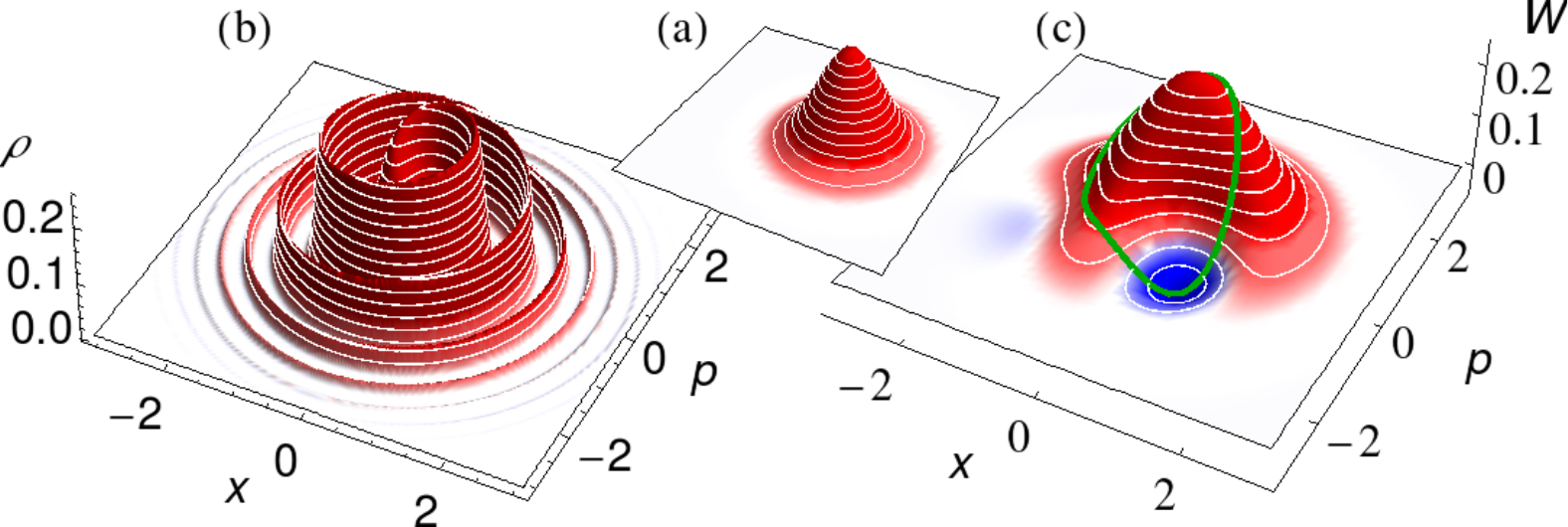}
  \caption{\emphCaption{Probability distributions in \ps.}\emphLabel{a},
    We start from a weakly excited coherent state~$|\alpha\rangle = | 7/12 \rangle$ which
    is positive everywhere.\emphLabel{b}, After time $t=60$ under classical time
    evolution, $\rho(t)$ has formed a highly sheared and tight spiral.\emphLabel{c}, After
    the same time the quantum evolution yields a Wigner distribution~$W(t)$ which has much
    less fine detail but nega\-tiv\-ities (blue). The green line superimposed
    on\emphLabel{c} traces out the Wigner distribution profile on a ring around the origin with
    radius~$r=1$ which passes through an area with pronounced
    nega\-tivity.  Fig.~\ref{fig:Evolution} displays evolution on this ring
    [$\Lambda =\frac{1}{2}$, see Eq.~(\ref{eq_Kerr_Hamiltonian})].
    \label{fig:Distributions}}
\end{figure}

We introduce Kerr oscillators, their Wigner distribution $W$, and their associated Wigner
current~$\VEC{J}$ in Sec.~\ref{sec_Kerr_osc}.  In Sec.~\ref{sec_No_Trajectories} we
show that there are no trajectories and no phase-space flow for anharmonic systems such as Kerr
oscillators. In Sec.~\ref{sec_Qu_pulses} we investigate how pulses in \ps smear out
classical spirals [Fig.~\ref{fig:Distributions}\emphLabel{b}; in all figures atomic units
with $\hbar =1$, $M=1$ and $k=1$ are used]. We find that pulses in \ps steepen and lengthen
dynamically. This analysis is aided by the system's circular symmetry and the fact that
the probability on circles in \ps is conserved.  In Sec.~\ref{sec_viscosity} we show
that using Wigner current~$\VEC{J}$'s effec\-tive ``viscosity"~\cite{Oliva_Shear_19} allows
us to contrast classical with quantum dynamics and pick out special quantum states.

Our results can be generalised to higher-dimensional systems~\cite{Wigner_PR32}.

\section{Wigner distributions and Wigner current of Kerr oscillators\label{sec_Kerr_osc}}

A one-dimensional system's Wigner distribution
$W_\varrho(x,p,t)$~\cite{Wigner_PR32,Hillery_PR84} (where $x$ denotes position, $p$ the
associated momentum, and $t$ time), for a quantum state described by a density
matrix~$\hat \varrho$, is defined as the Fourier transform of its off-diagonal
coherences~$\varrho(x+y,x-y,t)$ (parametrized by the shift~$y$)
\begin{eqnarray}\label{eq:W}
  W_\varrho(x,p,t) = \frac{1}{\pi \hbar} \int_{-\infty}^{\infty} dy
  \; \langle x+y | \hat
  \varrho(t) |x-y \rangle \; e^{-\frac{2i}{\hbar} p y},  
  \quad 
\end{eqnarray}
where $\hbar=h/(2\pi)$ is Planck's constant. By construction $W$ is normalized and
nonlocal (through~$y$). Unlike~$\hat \varrho$, $W$ is always real-valued but,
generically, $W$ features nega\-tivities~\cite{Wigner_PR32}. Since~$W_\varrho$
is~$\hat \varrho$'s Fourier transform, $W$ and~$\hat \varrho$ are isomorphic to each
other, allowing us to describe all aspects of the quantum system's state and its dynamics
using the Wigner representation of quantum theory~\cite{Zachos_book_05}.

\subsection{Time evolution of the Wigner distribution\label{sec_TimeEvol}}

For conservative Kerr systems the time development of~$W$ is given by the Moyal-bracket
$\{\{.,.\}\}$~\cite{Moyal_MPCPS49,Zachos_book_05}
\begin{flalign}
\label{eq:_MoyalProduct}
&\partial_t W(x,p,t)  =  \{\{H,W\}\}\ 
\\
\label{eq:_MoyalProductExplicit}
&\; \equiv \frac{2}{\hbar} ~ H(x,p)\ \sin \left (
  {{\tfrac{\hbar }{2}}(\stackrel{\leftarrow }{\partial }_x \stackrel{\rightarrow
    }{\partial }_{p}-\stackrel{\leftarrow }{\partial }_{p}\stackrel{\rightarrow }{\partial
    }_{x})} \right ) \ W(x,p,t).
\end{flalign}
Here, $\partial_x = \frac{\partial }{\partial x }$, etc.;
the arrows over the derivatives indicate whether they act on (point towards) Hamiltonian
or Wigner distribution.

The Hamiltonian of anharmonic single-mode oscillators of the Kerr type has the form
\begin{equation}
\label{eq_Kerr_Hamiltonian} {\hat H}_\Lambda = \left( \frac{\hat p^2}{2M} + \frac{k}{2} \hat
  x^2 \right) + \Lambda^2 \left(\frac{\hat p^2}{2M} + \frac{k}{2} \hat x^2 \right)^2 ,
\end{equation}
with the oscillator mass $M$ and spring constant $k$. Such Hamiltonians describe
electromagnetic fields subjected to Kerr nonlinearities~$\chi^{(3)}$ (here
$\Lambda^2 \propto
{\chi^{(3)}}$)~\cite{Walls_Milburn_QuopBook,Osborn_JPA09,Manko_PSc10,Kirchmair_NAT13}.
This system is fully solvable since wave functions of the harmonic oscillator are solutions
to the Kerr Hamiltonian with
eigenenergies~$E_n=\hbar \sqrt{\frac{k}{M}} [(n+\frac{1}{2}) + \Lambda^2
(n+\frac{1}{2})^2]$. Its quantum recurrence time is%~$
\begin{flalign} 
  T_\Lambda = \frac{\pi}{|\Lambda|^2}.
\end{flalign} 

Following Wigner~\cite{Wigner_PR32}, we cast expression~(\ref{eq:_MoyalProductExplicit}) in
the form of the phase-space continuity equation
\begin{flalign} 
&
\partial_t W +\bm \nabla \cdot {\VEC J} = 0 \; ,
\label{eq:W_Continuity}
\\ & \mbox{ where } \bm \nabla = \genfrac{(}{)}{0pt}{}{\partial_x}{\partial_p} \mbox{ is
  the gradient, and } {\VEC J} = \genfrac{(}{)}{0pt}{}{J_x}{J_p}  \nonumber
\end{flalign}
denotes the Wigner current in \ps~\cite{Ole_PRL13}. $\VEC{J}$ is the quantum
analog~\cite{Donoso_PRL01,Bauke_2011arXiv1101.2683B} of the classical phase-space
current~$\VEC{j} = \rho \VEC{v}$~\cite{Nolte_PT10} which transports the classical
probability density~$\rho(x,p,t)$ according to Liouville's continuity
equation~$\partial_t \rho = - {\bm \nabla} \cdot \VEC{j}$.

${\VEC J}$ reveals details~\cite{Ole_PRL13,Kakofengitis_EPJP17} about quantum systems'
phase space dynamics previously thought inaccessible due to the supposed ``blurring" by
Heisenberg's uncertainty principle.

\begin{widetext}
  From now on we will consider~$M=k=1$ only. Then [for a derivation see
  Eqs.~(\ref{eq:J_x}) and~(\ref{eq:J_p}) in the Appendix~\ref{sec:Appendix}],
  with~${\VEC r}=(x,p)=r(\cos \theta, \sin \theta)$, $r=\sqrt{x^2+p^2}$
  and~$\bm \Delta = \partial_x^2 + \partial_p^2$,~${\VEC J}$ can be written as
\begin{flalign}
  {\VEC J} & = \genfrac{(}{)}{0pt}{}{p}{-x}  
  \left[ 1+\Lambda^2 \left( x^2 +  p^2  -{ \frac {\hbar^2}{4}} \bm \Delta 
     \right) \right] W 
   = \genfrac{(}{)}{0pt}{}{r \sin \theta}{-r \cos \theta} \left[ 1 + \Lambda^2 \left( r^2
       -{ \frac {\hbar^2}{4}} (\partial_{r}^{2} +\frac{1}{r} \partial_{r} + \frac{1}{r^2}
       \partial_{\theta}^2 ) \right) \right] W \; .
\label{eq:J_polar} 
\end{flalign}
${\VEC J}$ is tangent to circles concentric with the origin of \ps. This \emph{circular
  symmetry} allows us to consider an approximation of the dynamics on such individual
circles, an observation we make use of below.
\end{widetext}

For future reference we split $\VEC{J}$ into its classical~$\VEC{j}$ and quantum
terms~$\VEC{J}^Q$
\begin{flalign} {\VEC{J}} & = \VEC{j} + \VEC{J}^Q
  = W  \VEC{v} -  \genfrac{(}{)}{0pt}{}{p}{-x}  
   \left({ \frac {\hbar^2 \Lambda^2}{4}} \bm \Delta 
     \right) W . \label{eq:CurrentComponents}
\end{flalign}
Here $\VEC{v}=\genfrac{(}{)}{0pt}{}{p}{-x}(1+\Lambda^2 r^2)$ is the classical phase-space
velocity. The quantum terms~$\VEC{J}^Q$ are only present for anharmonic
potentials~\cite{Kakofengitis_PRA17}, which is why only anharmonic potentials create
coherences. Harmonic systems' phase-space dynamics follows~$\VEC{v}$ and is classical, see
Refs.~\cite{Oliva_PhysA17,Kakofengitis_PRA17}.

\section{No trajectories or flow in quantum \ps\label{sec_No_Trajectories}}

Inspired by classical mechanics, there have been several attempts to treat quantum phase-space
evolution as a flow along trajectories~\cite{Oliva_PhysA17}. Such attempts are ill
fated~\cite{Oliva_PhysA17} as we explain now. They use the formal factorization
$\VEC J = W \VEC{w}$ to define a ``quantum phase-space velocity"~$\VEC{w} = \VEC J / W$, then the
continuity equation~(\ref{eq:W_Continuity}) assumes the
form~\cite{Trahan_JCP03,Daligault_PRA03,Oliva_PhysA17}
\begin{equation} \partial_t W + \VEC{w} \cdot \bm \nabla W + W \bm \nabla \cdot \VEC{w} = 0 \; .
\label{eq:W_ContinuityProdForm}
\end{equation}
Here the convective term~$\VEC{w} \cdot \bm \nabla W $ describes the transport that carries
$W$ along with the current (following fieldlines in \ps) without changing its values. In
contrast, the current divergence term~$W \bm \nabla \cdot \VEC{w}$ changes values
of~$W$. This is best seen by formally rearranging Eq.~(\ref{eq:W_ContinuityProdForm}) for
the total derivative
\begin{equation} 
  \frac{d W}{d t} = \partial_t W + \VEC{w} \cdot \bm \nabla W =- W \bm \nabla \cdot \VEC{w} \; . 
\label{eq:W_TotalDeriv}
\end{equation}
Treating a continuity equation in this form is known as its \emph{Lagrange
  decomposition}. This decomposition has to be treated with extreme caution, since it
essentially splits the well behaved and finite term $\bm \nabla \cdot \bm J$ into the two
individually singular terms~$\VEC{w} \cdot \bm \nabla W$ and~$ W \bm \nabla \cdot \bm
w$. Some implications are discussed below.

\begin{widetext}
For the Kerr system this total derivative is
\begin{flalign}
  \frac{d W}{d t} & = -\frac{\Lambda^2 \hbar^2 }{4} \left[ p \left( \frac{(\partial_x
        W)}{W} - \partial_x \right) \right.  \left. - x \left(\frac{(\partial_p W)}{W} -
      \partial_p \right)\right] \bm \Delta W = -\frac{\Lambda^2 \hbar^2 }{4} W
  \partial_\theta \left( \frac{ \bm \Delta W }{W} \right) ,
 \label{eq:W_div_w}        
\end{flalign}
\end{widetext}
and the convective transport term in Eq.~(\ref{eq:W_TotalDeriv}) is
\begin{flalign}
  \VEC{w} \cdot \bm \nabla W = \left( \Lambda^2 \left[-r^2 + \frac{\hbar^2}{4 W} \bm
      \Delta W \right] - 1 \right) \partial_\theta W .
  \label{eq:w_grad_W}        
\end{flalign}
Since the divergence~$ \bm \nabla \cdot \VEC{w}$ is nonzero, the quantum evolution does not
preserve phase-space volumes~\cite{Moyal_MPCPS49,Kakofengitis_PRA17,Oliva_PhysA17}. 

One could still describe quantum evolution by phase-space transport if the magnitude of
this divergence were finite across the entire phase space~\cite{Oliva_PhysA17}.  Indeed,
modelling quantum phase-space dynamics through such transport along trajectories has been
attempted many times; in this context it has been consi\-der\-ed an undesirable feature
of~$\VEC{w}$ that it is a singular quantity when~$W$ is zero (see
Ref.~\cite{Oliva_PhysA17} for details). But zeros in~$W$ are
unavoidable~\cite{Hudson_RMP74}:

The singularities in~$ \bm \nabla \cdot \VEC{w}$ are a fundamental and necessary feature to
create nega\-tive regions in~$W$ and thus to create quantum coherences. Such singularities
are not a flaw. A velocity field~$\VEC{w}$ with posi\-tive
divergence that is bounded from above, $B > \bm \nabla \cdot \VEC{w} > 0$, 
will by itself not be able to
generate nega\-tivities. 
The associated expansion of phase-space volumes can only reduce the
initial value~$W(0)>0$ of a density towards zero, 
since Eq.~(\ref{eq:W_TotalDeriv})
implies that~\cite{Trahan_JCP03,Oliva_PhysA17}
\begin{flalign}
  W(t)|_{\text {comoving}} > W(0) \exp(- B t) > 0
  \label{eq:W_drops_towards_zero}
\end{flalign}
for all times. Trahan and Wyatt noticed this and concluded that ``\emph{the sign of the
  density riding along the trajectory cannot change}''~\cite{Trahan_JCP03}.

But this interpretation is incorrect. When~$W=0$ the velocity $\VEC w$ and its divergence
is singular, Eq.~(\ref{eq:W_div_w}) cannot be integrated since
$\VEC w$'s singularities render integrals and associated bounds such
as~(\ref{eq:W_drops_towards_zero}) ill-defined~\cite{Oliva_PhysA17}. Therefore, in
anharmonic quantum systems neither trajectories nor transport along flow lines
exist~\cite{Oliva_PhysA17}. (Refs.~\cite{Bauke_2011arXiv1101.2683B}
and~\cite{Ole_PRL13} refer to Wigner ``flow" but were written before this 
was
realized.)

Because of the singular volume changes associated with Eq.~(\ref{eq:W_div_w}), we feel the
quantum \emph{Liouville} equation~(\ref{eq:W_Continuity}) should be called Wigner's
continuity equation instead.

We are forced to conclude that a trajectory-based approach to quantum phase-space evolution
creates contradictions such as singular~$\VEC{w}$ and singular phase-space volume changes. This
highlights the stark differences between classical and quantum dyna\-mics in an
illuminating manner.  The singularities in~$\VEC{w}$ and phase-space volume changes are needed to
violate inequality~(\ref{eq:W_drops_towards_zero}) thus allowing for the creation of
quantum coherences and nega\-tive regions in~$W$~\cite{Kakofengitis_PRA17,Oliva_PhysA17}.

\section{Pulses in Quantum \ps\label{sec_Qu_pulses}}

In the classical case the probability (of~$\rho$) on a classical trajectory of a
conservative system is conserved over time.  It can be checked that the probability
(of~$W$) on a classical trajectory is not conserved for typical anharmonic quantum
systems.

The quantum Kerr system is an exception as its evolution preserves probability on
rings around the origin:
\begin{flalign}
  \ointctrclockwise d\theta \;\; \partial_t W = - \ointctrclockwise d\theta \;\; \bm
  \nabla \cdot {\VEC J} =0 ,
  \label{eq:_W_ring_preserved}
\end{flalign}
since
$ \bm \nabla \cdot {\bm J} = r \partial_\theta([v(r) - \Lambda^2 \frac{\hbar^2}{4} \bm
\Delta]W) $.  In addition to the circular symmetry displayed in Eq.~(\ref{eq:J_polar}),
this probability conservation on circles is the primary reason why considering the Kerr
dynamics on circles is suitable.

\begin{figure}[b]
  \centering
  \includegraphics[width=0.48\textwidth]{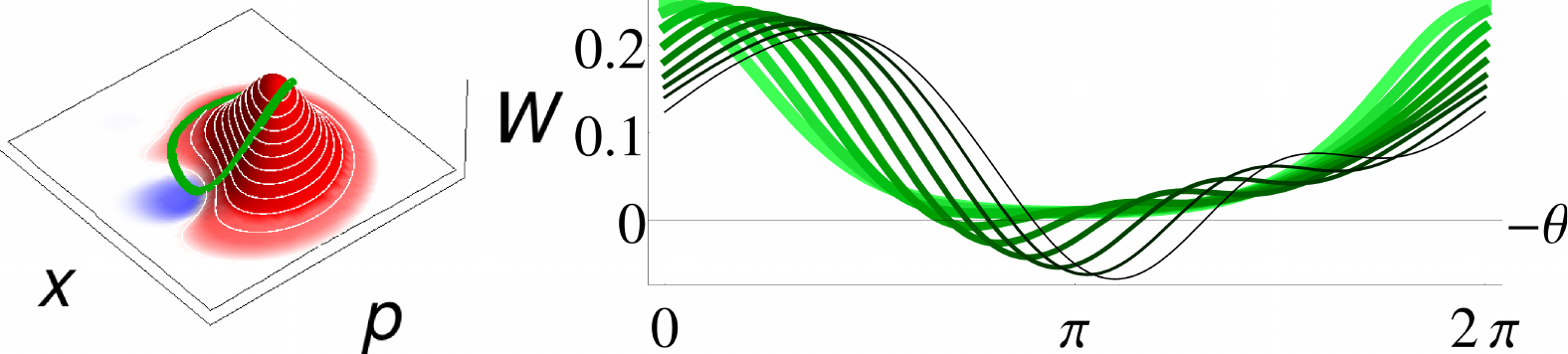}
  \caption{\emphCaption{Time evolution of $W(\theta)$ on a ring with fixed
      radius $r=1.0$} for initial coherent state~$|\alpha \rangle=| 7/12 \rangle$ over
    time $t=0$ to~$T_\Lambda /4 = 4 \pi = 12.56$ [$\Lambda = \frac{1}{4}$]. The darker
  and thinner the curves, the more time has elapsed. The curves move clockwise on the
  ring, towards increasing values of ``$-\theta$". The quantum evolution leads to a
  speedup over the classical evolution (the classical phase angle~$v t$ is
  subtracted). Additionally, under quantum evolution the pulse widens and steepens at the
  front, this triggers the formation of oscillations with nega\-tive regions in front of
  the pulse which eventually catch up with the main pulse from ``behind".
    \label{fig:Evolution}}
\end{figure}

The classical velocity profile~$v(r)$ leads to the formation of fine detail in the
classical evolution: in the case of a Gaussian initial state, the state becomes wrapped
into a single tightly wound spiral [see Fig.~\ref{fig:Distributions}\emphLabel{b}]. The
quantum evolution shows this tendency of spiral wrapping as well, but while the formation
of fine detail is suppressed through ``viscous'' behaviour (see Sec.~\ref{sec_viscosity}),
nega\-tivities of the Wigner distribution emerge. To study this in more detail,
consider~$W$ on a ring of radius~$r$, as displayed in Fig.~\ref{fig:Evolution}.

The quantum ``cross-talk" terms~$ \partial_{r}^{2} +\frac{1}{r} \partial_{r}$ in
Eq.~(\ref{eq:J_polar}) couple the current on adjacent rings. We can cast these terms aside
if we may assume that the Wigner distribution's azimuthal curvature
$ \partial_{\theta}^2 W $ is much greater than its radial curvature and gradient. Making
this assumption temporarily, the velocity on a ring is approximately
\begin{align}
  w(r,\theta) \approx r \left[ 1+\Lambda^2 \left( r^2 -
  { \frac {\hbar^2}{4 r^2}} \frac{1}{W} \partial_{\theta}^2 W \right) \right] .
\label{eq:v_ring}
\end{align}

This approximation is obviously poor when $W\approx0$, but Eq.~(\ref{eq:v_ring}) is still
useful for the discussion that follows.

\begin{figure}[t]
  \centering
  \includegraphics[width=0.48\textwidth]{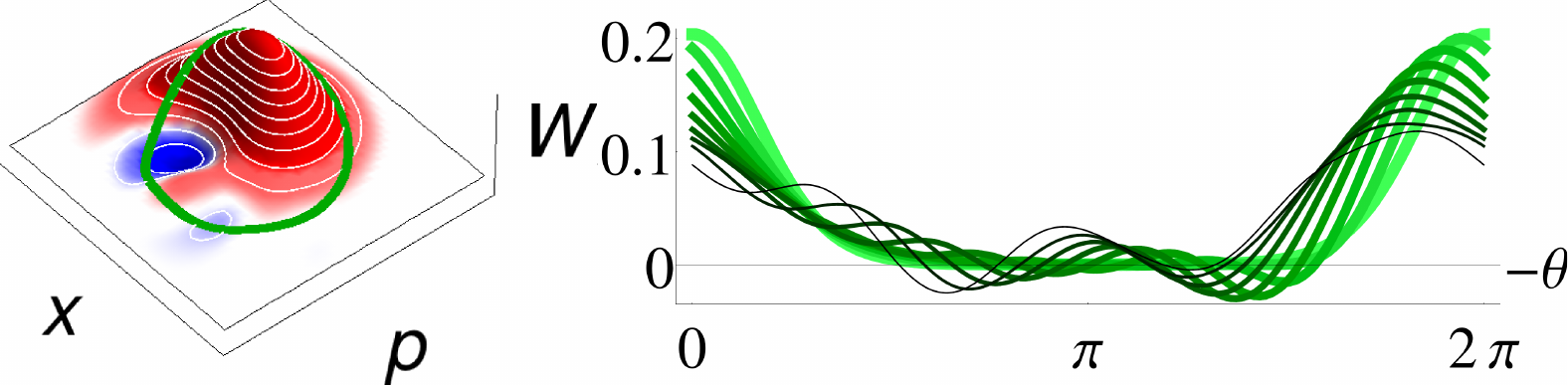}
  \caption{\emphCaption{Time evolution of $W(\theta)$ on a ring with fixed
      radius $r=1.6$} for initial coherent state~$|\alpha \rangle=| 5/4 \rangle$ over time
    $t=0$ to~$T_{\Lambda_-} /4 = 12.56$. Here the Kerr-nonlinearity is \emph{nega\-tive},
    $\Lambda_-^2 = - 1/16$,
    therefore the Wigner distribution is wrapped \emph{anticlockwise} and the
    center of gravity of the pulse falls \emph{behind} the classical motion ($v t$ has
    been subtracted). Contrast with Fig.~\ref{fig:Evolution}.
    \label{fig:HigherExcitationCoherentState}}
\end{figure}

\begin{figure}[t]
  \centering
  \includegraphics[width=0.48\textwidth]{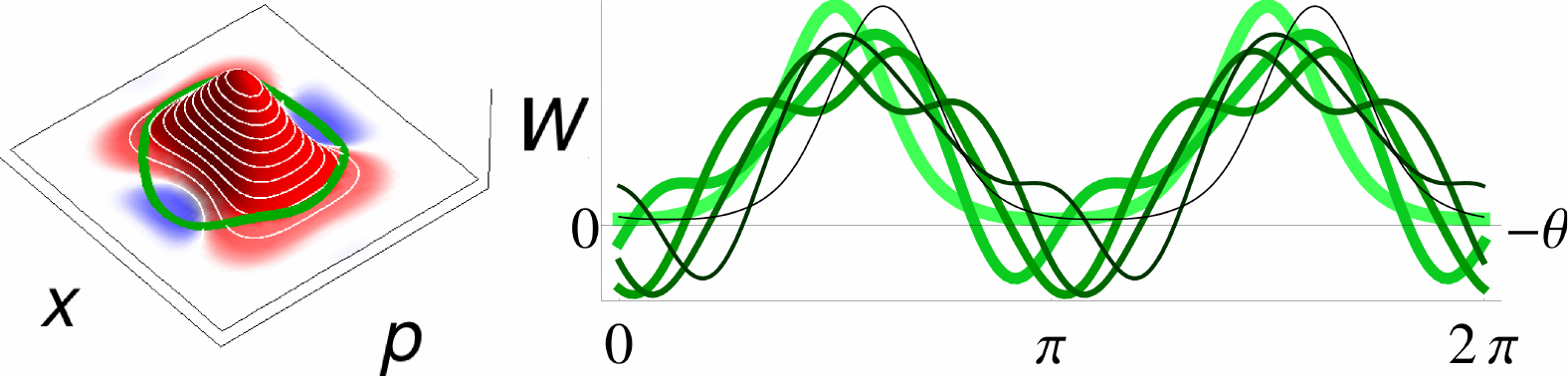}
  \caption{\emphCaption{Time evolution of $W(\theta)$ on a ring with fixed radius $r=1.6$}
    for initial squeezed vacuum state (squeezing parameter~$=1/3$) over time $t=0$ to
    $T_{\Lambda_+} = \frac{\pi}{4} = 0.785$.  Here the Kerr-nonlinearity is
    \emph{positive}, $\Lambda_+^2 = 1$, therefore the Wigner distribution's
    center of gravity moves \emph{ahead} of the classical motion ($v t$ has been
    subtracted): at time~$t=\frac{T_1}{4}$ (its recurrence time is shortened
    by~$\frac{T_1}{4}$ because the squeezed state is symmetric with respect to the origin)
    the original pulse reforms and is rotated forward, similar to
    Fig.~\ref{fig:Evolution}.
    \label{fig:Wrings_SqueezedState}}
\end{figure}

In Figs.~\ref{fig:Evolution}-\ref{fig:Wrings_SqueezedState} the full 
evolution is
portrayed, not its approximate behaviour of Eq.~(\ref{eq:v_ring}).  
The axis ``$-\theta$"
is chosen in Figs.~\ref{fig:Evolution}-\ref{fig:Wrings_SqueezedState}
since classical evolution proceeds clockwise, in the direction of 
nega\-tive values of
$\theta$.

The effect of the $\theta$-curvature term, retained in Eq.~(\ref{eq:v_ring}), is primarily
twofold: for a Wigner distribution on a circle, forming a hump, the hump's leading and
trailing edges, having positive curvature, get delayed. Conversely, the nega\-tive
curvature of the peak of the hump accelerates its center (see
Fig.~\ref{fig:Evolution}). This lengthens the pulse, making the tail trail, and sharpens
its front since the center catches up with the front (see
Fig.~\ref{fig:Evolution}).  This
sharpening in turn spawns oscillations that project forward from the pulse (see
Fig.~\ref{fig:Evolution} and discussion in Ref.~\cite{Friedman_17}).

A narrower pulse, as portrayed in Fig.~\ref{fig:HigherExcitationCoherentState}, develops
more pronounced oscillations. Additionally, in
Fig.~\ref{fig:HigherExcitationCoherentState}, $\Lambda$ is chosen formally complex such
that $\Lambda^2_- < 0$. This creates ``backwards'' dynamics when contrasted with a
positive Kerr-nonlinearity (compare Figs.~\ref{fig:Evolution}
and~\ref{fig:HigherExcitationCoherentState}: in
Fig.~\ref{fig:HigherExcitationCoherentState} the pulse lengths to the ``right" and
steepens and spawns oscillations to the ``left"; in ``reverse" to
Fig.~\ref{fig:Evolution}).

In Fig.~\ref{fig:Wrings_SqueezedState}, two pulses on a ring interfere with each
other. Here, like in Fig.~\ref{fig:Evolution}, the overall effect is that the quantum
terms speed the pulses up.


\section{$\VEC J$'s viscosity and special states\label{sec_viscosity}}

In Sec.~\ref{sec_Qu_pulses} we discussed motion on a ring. Here we consider cross talk
between motion on neighbouring rings.

Over time classical Hamiltonian phase-space flow shears~$\rho$ since~$\VEC{v}$ creates nonzero
gradients of its angular velocity across energy shells. This flow is inviscid as $\VEC{v}$ is
independent of~$\rho$; thus no terms suppress the effects of the angular velocity
gradients, and, as time progresses, nonsingular probability distributions in \ps get sheared
into ever finer filaments [see~Fig.~\ref{fig:Distributions}\emphLabel{b}].

The associated classical phase-space shear has been derived in Ref.~\cite{Oliva_Shear_19} as
\begin{flalign}
  s (x,p;H) = \partial_{\widehat {{\bm \nabla}}_{\!\!H}} ( - {\bm \nabla} \times  \VEC{v} ) =
\partial_{\widehat {{\bm \nabla}}_{\!\!H}} (  \partial_p v_x - \partial_x v_p) . 
\label{eq:_ClassicalShear}
\end{flalign}
Here the directional derivative across energy
shells,~$\partial_{\widehat {{\bm \nabla}}_{\!\!H}}$, is formed from the normalized
gradient $\widehat {\bm \nabla}_{\!\!H} = {\bm \nabla} H/|{\bm \nabla} H|$ of the
Hamiltonian~$H$. Because of the Kerr system's circular
symmetry,~$\widehat {\bm \nabla}_{\!\!H} = \partial_r$.

The sign convention using the nega\-tive curl in~$s$ in Eq.~(\ref{eq:_ClassicalShear}) is
designed to yield a positive sign for \emph{clockwise}-orientated fields since this is the
prevailing direction of the classical velocity field~$\VEC{v}$.  This choice yields $s>0$
for hard potentials (potentials for which the magnitude of the force increases with
increasing amplitude, i.e., $\Lambda^2>0$), since they induce \emph{clockwise} shear
[see~Fig.~\ref{fig:Distributions}\emphLabel{b}]. $s=0$ for harmonic oscillators (i.e.,
$\Lambda = 0$), and $s<0$ for soft potentials (for which the magnitude of the
force decreases with increasing amplitude, i.e., $\Lambda^2<0$) since they induce
\emph{anticlockwise} shear.
\begin{figure}[t]
\includegraphics[width=1.05\columnwidth]{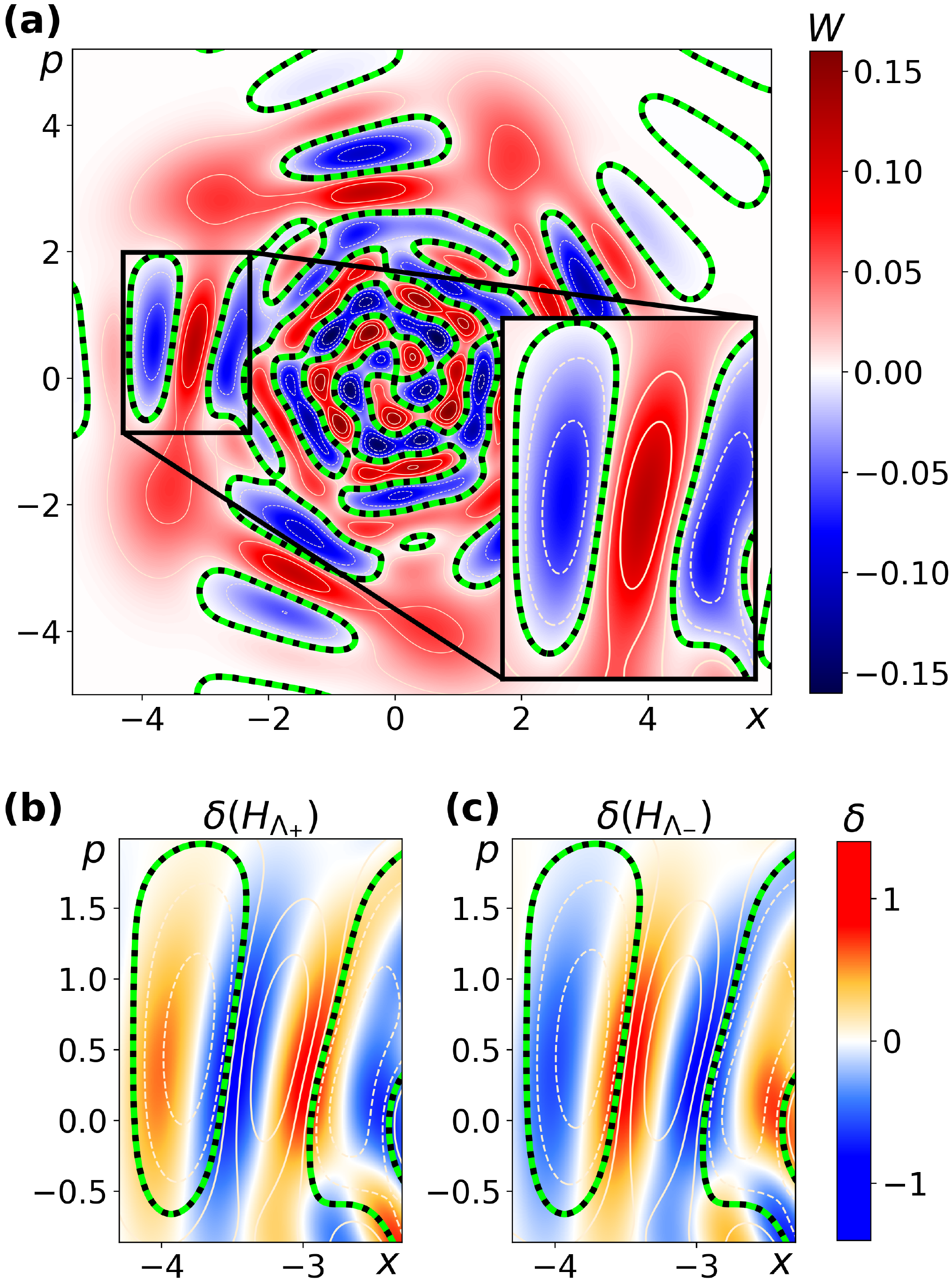}
\caption{\CO{(Colour online)} \emphCaption{Polarization of the vorticity $\delta$ and
    inversion of this polarization.}  \emphLabel{a}, The Wigner distribution~$W$ of a
  Gaussian initial state centered on $x=-4$, $p=0$ and evolved to $t=40$ using
  $\hat H_{\Lambda_+} = \hat H_{{1}/{4}}$. Its contours (for emphasis the zero contour is
  shown as black-green dashed lines) are also employed in \emphLabel{b} and \emphLabel{c}.
  The inset for $W$ in \emphLabel{a} is reproduced showing the effects of,\emphLabel{b},
  clockwise shear [$\delta(H_{\Lambda_+})$] and,\emphLabel{c}, anticlockwise shear
  [$\delta(H_{\Lambda_-})$].  Comparing \emphLabel{b} with \emphLabel{c} demonstrates
  polarization inversion of $\delta$ associated with shear inversion of the system, here
  $\Lambda^2_+ = +(1/4)^2 = - \Lambda^2_-$.
 \label{fig:LocalShear}}
\end{figure}
The reaction of quantum dynamics to classical shear~$s$ has to reside in $\VEC{J}^Q$ of
Eq.~(\ref{eq:CurrentComponents}). To extract it we form the vorticity
of~$\VEC{J}^Q$~\cite{Oliva_Shear_19}:
\begin{flalign}
  \delta (x,p,t;H) = - {\bm \nabla} \times  \VEC{J}^Q = \partial_p J_x^Q - \partial_x J_p^Q \; . 
\label{eq:shear_deviation}
\end{flalign}
$\delta$'s sign distribution shows a pronounced polarization pattern, see
Fig.~\ref{fig:LocalShear}.

Specifically, for a system with clockwise shear Fig.~\ref{fig:LocalShear}\emphLabel{b}
illustrates that $\delta(H_{\Lambda_+})$ [with $\Lambda^2_+ = + (1/4)^2$] tends to be
posi\-tive on the inside (towards the origin) and nega\-tive on the outside of the
positive main ridge of~$W$ [see inset of Fig.~\ref{fig:LocalShear}\emphLabel{a}].
Because of this, the outside is being slowed down while the inside speeds up. This
polarized distribution of~$\delta$ therefore counteracts the classical
shear~($s_{H_{\Lambda_+}} > 0$) and can suppress it altogether~\cite{Oliva_Shear_19}. The
same applies to other positive regions of~$W$, whereas for its nega\-tive regions the
current~$\VEC{J}$ tends to be inverted~\cite{Ole_PRL13,Kakofengitis_EPJP17},
inverting~$\delta$'s polarization pattern [see Ref.~\cite{Oliva_Shear_19} and
Fig.~\ref{fig:LocalShear}\emphLabel{b}].

When the same state~$W$ is governed by a Hamiltonian $H_{\Lambda_-}$ with anticlockwise
shear~\cite{Oliva_Shear_19} [i.e., $(\Lambda_-)^2<0$], $\delta(H_{\Lambda_-})$ tends to be
the sign-inverted form of~$\delta(H_{\Lambda_{+}})$~(for Kerr systems we find
$\delta(H_{\Lambda_{+}}) = - \delta(H_{\Lambda_{-}})$ if $|\Lambda_{+}|=|\Lambda_{-}|$).
This is illustrated in Fig.~\ref{fig:LocalShear}\emphLabel{c}, where
$\Lambda^2_- = - (1/4)^2$ is negative, whereas in Fig.~\ref{fig:LocalShear}\emphLabel{b}
$\Lambda^2_+ = +(1/4)^2$ is positive.

The distribution of~$\delta$'s polarization can be picked up with the directional
derivative~$\partial_{\widehat {{\bm \nabla}}_{\!\!H}} \delta(t;H) = \partial_{r}
\delta(t;H)$.  This we multiply with~$W$, because nega\-tive regions of~$W$ invert the
current~$\VEC{J}$~\cite{Ole_PRL13}, and because we want to weight it with the local
contribution of the state. The resulting local measure for weighted shear polarization
is~\cite{Oliva_Shear_19} ${\pi}(x,p,t;H) = W(t) \; \partial_{r} \delta(t;H)$.  Its average
across \ps is $W$'s shear polarization~\cite{Oliva_Shear_19}
\begin{flalign}
  {\Pi}(t;H) = \langle \!\!\,\langle {\pi}(t;H) \rangle \!\!\, \rangle =
  \iint_{-\infty}^{\infty} dx dp \; \pi(x,p,t;H) \, .
\label{eq:effective_viscosity_quantification}
\end{flalign}

\begin{figure}[b]
  \centering
  \includegraphics[width=0.99\columnwidth,height=0.55\columnwidth]{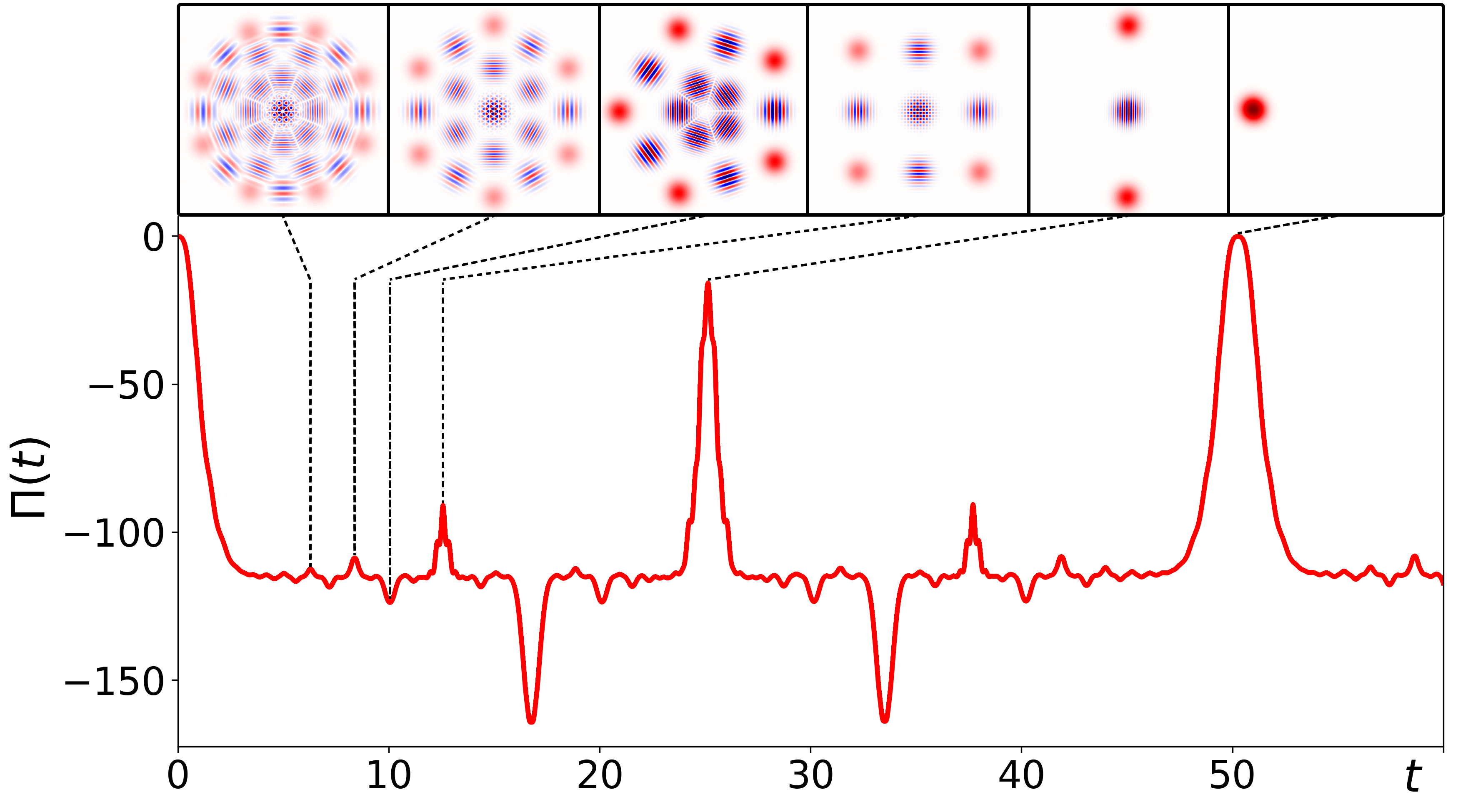}
  \caption{\emphCaption{Smoothed $\Pi(t)$ picks out special states.} Deviations of ${\Pi}(t)$ from
    the settled value ($\approx -115$) single out special states: the evolution shows recurrence of the
    initial state at time~$T_{\Lambda_+}= 16 \pi \approx 50.3$ ($\Lambda_+ = 1/2$). Pronounced peaks and
    troughs at intermediate
    times identify fractional revival states~\cite{Averbukh_PLA89} with special $n$-fold symmetries.
    \label{fig:GlobalShear}}
\end{figure}

Fig.~\ref{fig:GlobalShear} illustrates that $\Pi(t)$ initially drops and after a while
levels off.

We emphasize that the levelling-off behaviour of~$\Pi(t)$ is in marked contrast to the
classical case: for long enough times, in simple bound-state classical systems
nonsingu\-lar states~$\rho(t)$ get stretched out linearly~\cite{Oliva_Shear_19} into ever
finer threads [see Fig.~\ref{fig:Distributions}\emphLabel{b}] therefore
$ \langle \!\!\,\langle \partial_{r} (- \VEC{\nabla} \times \VEC{j}) \rangle \!\!\,
\rangle \propto t$~\cite{Oliva_Shear_19}. The quantum evolution counteracts this classical
shear $s$ resulting in values of the shear suppression $\Pi$ which are opposite in sign to
those of $s$~\cite{Oliva_Shear_19} (for the Kerr system sgn[$s$]=sgn[$\Lambda^2$]).  

Moreover, starting from an initial Gaussian state, the magnitude $|\Pi(t)|$ initially
grows the more the evolution stretches out the state into finer structures. Eventually
quantum shear suppression stops classical shear from creating finer structures in
\ps~\cite{Oliva_Shear_19}: $|\Pi(t)|$ levels off.

In other words, the quantum evolution is effectively ``viscous".~This viscosity is the
mechanism by which quantum evolution enforces that~$W$ can typically not form structures
below the size scale identified by Zurek~\cite{Zurek_NAT01}.~Therefore, $\Pi(t)$ settles
when the state has formed structures at the Zurek scale.~This can,~e.g.,~be quantified by
monitoring the phase-spatial frequency content of $W$ as a function of time (for details
see~Ref.\cite{Oliva_Shear_19}).

Yet, quantum evolution is not truly viscous, it allows for revivals. Interestingly, these
are picked up by the \emph{deviation} of $\Pi(t)$ from the local time average. For the
Kerr system, the special states for which this deviation is largest are (fractional)
revival states~\cite{Robinett_PR04,Kirchmair_NAT13} (see Fig.~\ref{fig:GlobalShear}).  

We emphasize that such revival states are traditionally picked up through the overlap of
the evolved state with a suitably chosen reference state (such as a Gaussian initial
state)~\cite{Robinett_PR04}, instead, our measure $\Pi(t)$ does not depend on a reference
state, this makes it more versatile than the use of wave-function overlaps.

We note that graphs of $\Pi(t)$ for anharmonic systems that do not have the symmetry of
the Kerr system carry high frequency oscillations~\cite{Oliva_Shear_19}, whereas, due to
the symmetry of the Kerr system, such oscillations are absent here. Generally, for other
anharmonic systems without circular symmetry, graphs of as smooth as those for $\Pi(t)$
obtained in Fig.~\ref{fig:GlobalShear} require frequency filtering~\cite{Oliva_Shear_19}.
In addition to the symmetries identified above, also in this regard are Kerr oscillators
the simplest possible continuous quantum systems that alter quantum coherences.

\emph{To conclude,} quantum dynamics that generates coherences in continuous systems is
most easily studied in \ps and using Kerr systems, since these have special symmetries.
The two symmetries we have identified are circular phase-space current~$\VEC J$,
Eq.~(\ref{eq:J_polar}), and probability conservation for~$W$ on rings,
Eq.~(\ref{eq:_W_ring_preserved}). These imply the absence of high-frequency components in
$\Pi(t)$ of Eq.~(\ref{eq:effective_viscosity_quantification}), see
Fig.~\ref{fig:GlobalShear}. We also have identified a quantum speedup of the propagation
of wave-function pulses in \ps and we demonstrate that the dynamics of the Kerr system is
``effectively viscous".  This can be quantified, explains the emergence of Zurek's scale
for the formation of minimum structures in quantum \ps, and can be used to pick out
special quantum states.

The geometric nature of our approach helps us to guide the
understanding of the generation of coherences in quantum dynamics and the formation of
nega\-tivities of~$W$ and will hopefully help pave the way to devise new strategies to
protect coherences (for related ideas see Ref.~\cite{Friedman_17}).

\begin{acknowledgments}
  O.S. wants to thank Eran Ginossar for his suggestion to investigate the Kerr system and Paul
  Brumer for his encouragement to study the formation of nega\-tivities in \ps.
\end{acknowledgments}

%Email: \href{mailto:Ole.Steuernagel@gmail.com}{Ole.Steuernagel@gmail.com}

%\begin{widetext}
\onecolumngrid
\section{Appendix \label{sec:Appendix}}

The Hamiltonian of anharmonic single-mode oscillators of the Kerr type has the form~(\ref{eq_Kerr_Hamiltonian})
\begin{equation}
\label{eq_Kerr_HamiltonianAppendix} {H}_\Lambda = \left( \frac{ p^2}{2M} + \frac{k}{2} 
  x^2 \right) + \left(\Lambda \frac{ p^2}{2M} + \lambda \frac{k}{2}  x^2 \right)^2 ,
\end{equation}
with $\Lambda = \lambda$. Here we keep the two parameters $\Lambda$ and $\lambda$ distinct
to allow us to tune the system's nonlinearities independently and help with keeping track
of terms in the derivation of the form of~$\VEC J$.

The Wigner distribution of the Kerr oscillator obeys the phase-space continuity
equation~(\ref{eq:_MoyalProductExplicit})~\cite{Stobinska_PRA08,Manko_PSc10,Kelly_JCP12}
\begin{eqnarray}
\label{eq:_MoyalProduct2}
\partial_t W(x,p,t) & = & \{\{H,W\}\}\ 
= \frac{2}{\hbar} ~ H(x,p)\ \sin \left (
  {{\tfrac{\hbar }{2}}(\stackrel{\leftarrow }{\partial }_x \stackrel{\rightarrow
    }{\partial }_{p}-\stackrel{\leftarrow }{\partial }_{p}\stackrel{\rightarrow }{\partial
    }_{x})} \right ) \ W(x,p,t) 
\\
 & = & \left( \left[ -\Lambda^2 \frac{\hbar^2 }{4 M^2}p \partial_x^3 + \lambda^2
    \frac{\hbar^2 k^2}{4} x \partial_p^{3} - \left\{ \Lambda \lambda {\frac
        {kx{p}^{2}}{M}}+ \lambda^2 {k} ^{2}{x}^{3} \right\} \partial_p \right.
\right. \nonumber \\
& & \left.  \left. - \Lambda \lambda
    \frac {\hbar^2 k }{4M} p \partial_x \partial_p^2 + \Lambda \lambda \frac {\hbar^2 k
      }{4} x \partial_p \partial_x^2  + \left\{ \Lambda^2 {\frac {{p}^{3}}{{M}^{2}}}+ \Lambda \lambda {\frac
        {k{x}^{2}p}{M}} \right\} \partial_x \right] + \frac {p}{M} \partial_x - k
  x \partial_p \right) W (x,p,t) . \label{eq:WF_Kerr_converted2} 
\end{eqnarray}
The square brackets enclose the terms arising from the Kerr Hamiltonian's anharmonic part,
whereas the terms $ \frac{p}{M} \partial_x - k x \partial_p$ stem from the harmonic
oscillator contribution $p^2/(2M) + k x^2/2$.

The associated Wigner current components~(\ref{eq:W_Continuity}) are 
\begin{eqnarray}
  J_x & = & \left[\hbar^2 \left( -{ \Lambda^2\frac {1}{4{M}^{2}}} p \partial_{x}^{2} -
    \Lambda \lambda {\frac {k}{4 M}} p \partial_{p}^ {2} \right) + \left\{ { \Lambda^2
      \frac {{p}^{3}}{{M}^{2}}}+ \Lambda \lambda {\frac {k{x}^{2}p}{M}} +{\frac {p }{M}}
    \right\} \right] W(x,p,t)
\label{eq:J_x} 
\\ \mbox{and } J_p & = &\left[ \hbar^2 \left( \lambda^2 \frac{ k^2}{4} x \partial_{p}^2 + \Lambda
    \lambda \frac {k}{4 M} x \partial_{x}^2 \right) - \left\{
    \lambda^2 {k}^{2}{x}^{3} +\Lambda \lambda {\frac {kx{p }^{2}}{M}} + k x \right\}  \right] W(x,p,t) . 
\label{eq:J_p}
\end{eqnarray}

The curly brackets in Eqs.~(\ref{eq:J_x}) and~(\ref{eq:J_p}) contain the classical
Hamiltonian current terms, and the round brackets contain the quantum terms.

To justify this assignment, note that the first term in $J_p$ is of the
form~$\frac{\hbar^2}{4\cdot 3!} \partial_x^3 V \partial_p^2 W$~\cite{Wigner_PR32,Ole_PRL13}
and thus has to be assigned to $J_p$, while the first term of $J_x$ is its ``partner" term
for the position case. What remains somewhat ambiguous is whether the second terms
in~(\ref{eq:J_x}) and~(\ref{eq:J_p}) have been assigned correctly. To highlight this
ambiguity consider
\begin{eqnarray}
  J_x^{(\sigma)} & = & J_x + \sigma \Lambda \lambda  \frac{\hbar^2 k}{4 M}
                  \left[ x \partial_{p} \partial_x  + p \partial_p^2 \right] W(x,p,t) 
\label{eq:J_x_s} 
\\
\mbox{and }
J_p^{(\sigma)} & = & J_p - \sigma \Lambda \lambda \frac {\hbar^2 k}{4 M}
\left[ x \partial_x^2 + p \partial_{x} \partial_p \right] W(x,p,t) , 
\label{eq:J_p_s} 
\end{eqnarray}
parametrized by the interpolation parameter~$\sigma$ with $0 \leq \sigma \leq 1$. This
interpolation fulfils the continuity equation~(\ref{eq:W_Continuity}) since the
$\sigma$-dependent terms are divergence-free for~ $0 \leq \sigma \leq 1$. 

\begin{figure}[h]
  \centering
  \includegraphics[width=0.78\textwidth]{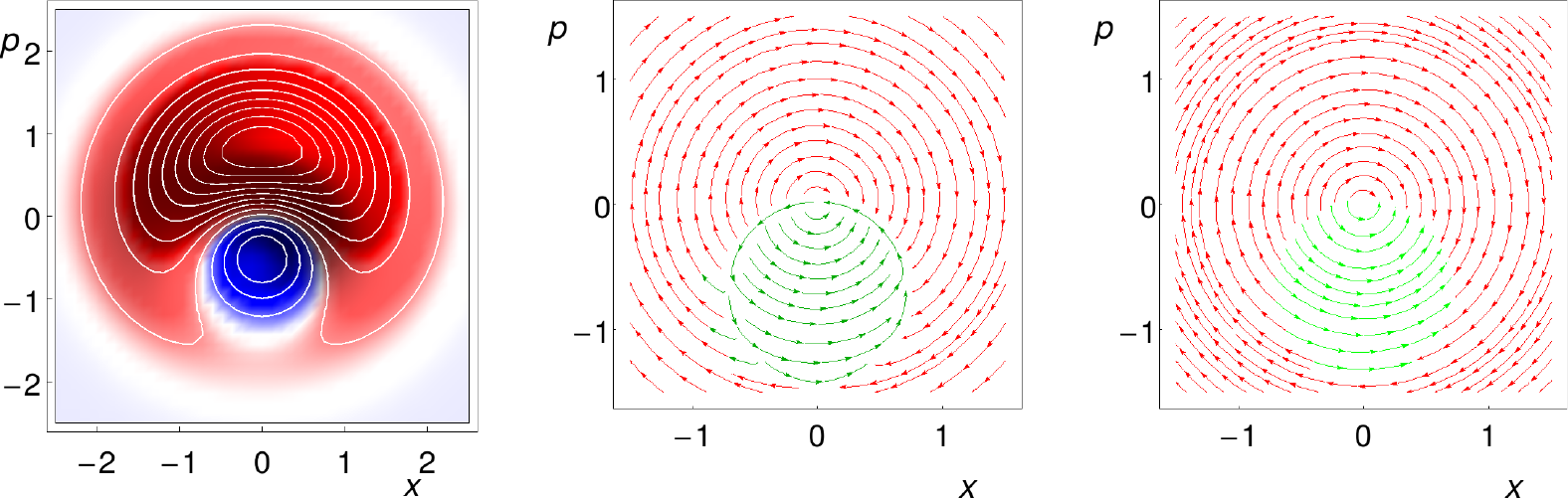}
  \caption{\emphCaption{Wigner distribution, incorrect and correct Wigner
      current patterns for state~$(|0\rangle +|1\rangle)/\sqrt{2}$.}  With
    $\Lambda = \lambda$ the dynamics of this superposition state is isomorphic to that of
    the harmonic oscillator, except for an extra phase due to the Kerr oscillator's
    different energy spectrum. The incorrect expression~$\VEC{J}^{(\sigma=1)}$ for the
    current (middle panel) does not respect this isomorphism; it breaks the system's
    circular symmetry and is therefore discarded. The correct
    expression~$\VEC{J}^{(\sigma=0)}$ for the current is depicted in the right-hand panel. The
    region represented by green is that where $W<0$; this leads to current
    inversion~\cite{Ole_PRL13}. For the Kerr system the only {point of
      stagnation}~\cite{Ole_PRL13} of the current is the coordinate origin. When the
    current stagnates elsewhere in \ps, it forms \emph{lines of
      stagnation}~\cite{Kakofengitis_EPJP17}.
    \label{fig_Appendix:current}}
\end{figure}

To remove the ambiguity we can use Wigner current plots. We notice that the field plots
of~$\VEC J^{(\sigma\neq 0)}$ do not ``make sense" [see Fig.~\ref{fig_Appendix:current}:
$\VEC{J}^{(\sigma=0)}$ of Eqs.~(\ref{eq:J_x}) and (\ref{eq:J_p}), or~Eq.~(\ref{eq:J_polar}) is
the correct Wigner current expression].

We emphasize that this circular symmetry of~$\VEC J$, derived for $W$ formed from a
superposition of two states, carries over to the case of general~$W$ since any $W$ can be
decomposed into sums of two-state superpositions.

%\end{widetext}
\twocolumngrid

%\onecolumngrid
%\newpage
%\input{Reply}

\end{document}